\documentclass[twoside,twocolumn,9pt]{article}
\usepackage{extsizes}
\usepackage[super,sort&compress,comma]{natbib} 
\usepackage[version=3]{mhchem}
\usepackage[left=1.5cm, right=1.5cm, top=1.785cm, bottom=2.0cm]{geometry}
\usepackage{balance}
\usepackage{mathptmx}
\usepackage{sectsty}
\usepackage{graphicx} 
\usepackage{lastpage}
\usepackage[format=plain,justification=justified,singlelinecheck=false,font={stretch=1.125,small,sf},labelfont=bf,labelsep=space]{caption}
\usepackage{float}
\usepackage{fancyhdr}
\usepackage{fnpos}
\usepackage[english]{babel}
\addto{\captionsenglish}{%
  
}
\usepackage{array}
\usepackage{droidsans}
\usepackage{charter}
\usepackage[T1]{fontenc}
\usepackage[usenames,dvipsnames]{xcolor}
\usepackage{setspace}
\usepackage[compact]{titlesec}

\usepackage{epstopdf}

\definecolor{cream}{RGB}{222,217,201}

\usepackage{changes}
\definechangesauthor[name={Indrajit's remarks}, color=blue]{IT}
\definechangesauthor[name={Tristan's remarks}, color=violet]{TS}
\definechangesauthor[name={Daniel's remarks}, color=orange]{DMS}

\usepackage{amssymb,amsmath} 
\usepackage{hyperref}
\begin{document}

\pagestyle{fancy}
\thispagestyle{plain}
\fancypagestyle{plain}{
\renewcommand{\headrulewidth}{0pt}
}

\makeFNbottom
\makeatletter
\renewcommand\LARGE{\@setfontsize\LARGE{15pt}{17}}
\renewcommand\Large{\@setfontsize\Large{12pt}{14}}
\renewcommand\large{\@setfontsize\large{10pt}{12}}
\renewcommand\footnotesize{\@setfontsize\footnotesize{7pt}{10}}
\makeatother

\renewcommand{\thefootnote}{\fnsymbol{footnote}}
\renewcommand\footnoterule{\vspace*{1pt}%
\color{cream}\hrule width 3.5in height 0.4pt \color{black}\vspace*{5pt}} 
\setcounter{secnumdepth}{5}

\makeatletter 
\renewcommand\@biblabel[1]{#1}            
\renewcommand\@makefntext[1]%
{\noindent\makebox[0pt][r]{\@thefnmark\,}#1}
\makeatother 
\renewcommand{\figurename}{\small{Fig.}~}
\sectionfont{\sffamily\Large}
\subsectionfont{\normalsize}
\subsubsectionfont{\bf}
\setstretch{1.125} 
\setlength{\skip\footins}{0.8cm}
\setlength{\footnotesep}{0.25cm}
\setlength{\jot}{10pt}
\titlespacing*{\section}{0pt}{4pt}{4pt}
\titlespacing*{\subsection}{0pt}{15pt}{1pt}

\fancyfoot{}
\fancyfoot[LO,RE]{\vspace{-7.1pt}\includegraphics[height=9pt]{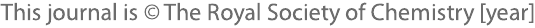}}
\fancyfoot[CO]{\vspace{-7.1pt}\hspace{13.2cm}\includegraphics{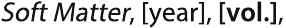}}
\fancyfoot[CE]{\vspace{-7.2pt}\hspace{-14.2cm}\includegraphics{head_foot/RF}}
\fancyfoot[RO]{\footnotesize{\sffamily{1--\pageref{LastPage} ~\textbar  \hspace{2pt}\thepage}}}
\fancyfoot[LE]{\footnotesize{\sffamily{\thepage~\textbar\hspace{3.45cm} 1--\pageref{LastPage}}}}
\fancyhead{}
\renewcommand{\headrulewidth}{0pt} 
\renewcommand{\footrulewidth}{0pt}
\setlength{\arrayrulewidth}{1pt}
\setlength{\columnsep}{6.5mm}
\setlength\bibsep{1pt}

\makeatletter 
\newlength{\figrulesep} 
\setlength{\figrulesep}{0.5\textfloatsep} 

\newcommand{\topfigrule}{\vspace*{-1pt}%
\noindent{\color{cream}\rule[-\figrulesep]{\columnwidth}{1.5pt}} }

\newcommand{\botfigrule}{\vspace*{-2pt}%
\noindent{\color{cream}\rule[\figrulesep]{\columnwidth}{1.5pt}} }

\newcommand{\dblfigrule}{\vspace*{-1pt}%
\noindent{\color{cream}\rule[-\figrulesep]{\textwidth}{1.5pt}} }

\makeatother

\twocolumn[
  \begin{@twocolumnfalse}
{\includegraphics[height=30pt]{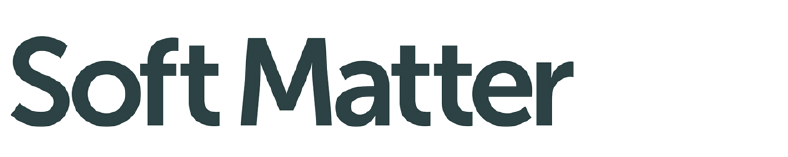}\hfill\raisebox{0pt}[0pt][0pt]{\includegraphics[height=55pt]{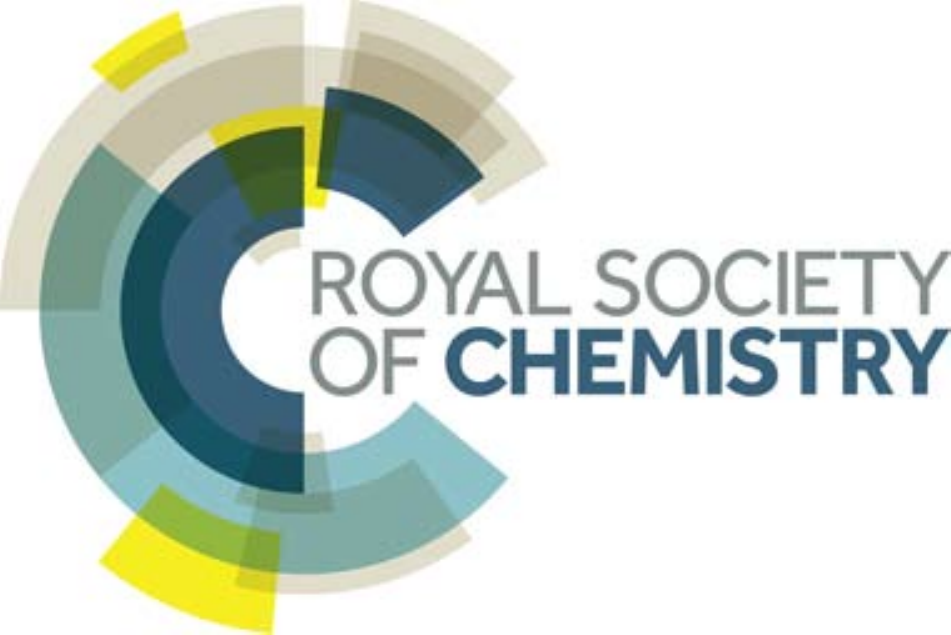}}\\[1ex]
\includegraphics[width=18.5cm]{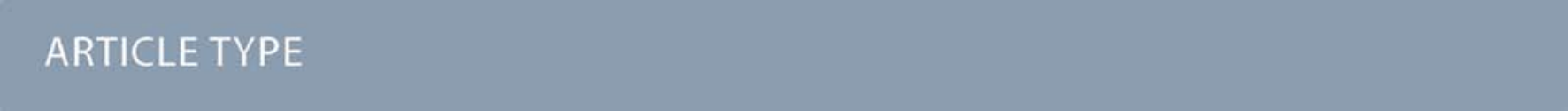}}\par
\vspace{1em}
\sffamily
\begin{tabular}{m{4.5cm} p{13.5cm} }

\includegraphics{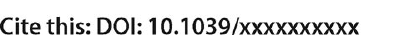} & \noindent\LARGE{\textbf{
Quantifying the link between local structure and cellular rearrangements using information in models of biological tissues}}\\
\vspace{0.3cm} & \vspace{0.3cm} \\

 & \noindent\large{ Indrajit Tah,\textit{$^{a}$} Tristan A. Sharp,\textit{$^{a}$} Andrea J. Liu,\textit{$^{a}$} and Daniel M. Sussman\textit{$^{b}$}} \\

\includegraphics{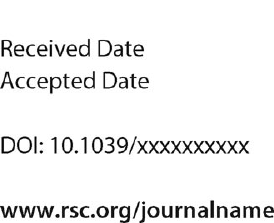} & \noindent\normalsize{
Machine learning techniques have been used to quantify the relationship between local structural features and variations in local dynamical activity in disordered glass-forming materials. To date these methods have been applied to an array of standard (Arrhenius and super-Arrhenius) glass formers, where work on ``soft spots'' indicates a connection between the linear vibrational response of a configuration and the energy barriers to non-linear deformations. Here we study the Voronoi model, which takes its inspiration from dense epithelial monolayers and which displays anomalous, sub-Arrhenius scaling of its dynamical relaxation time with decreasing temperature. Despite these differences, we find that the likelihood of rearrangements can vary by several orders of magnitude within the model tissue and extract a local structural quantity, ``softness" that accurately predicts the temperature-dependence of the relaxation time. We use an information-theoretic measure to quantify the extent to which softness determines impending topological rearrangements; we find that softness captures nearly all of the information about rearrangements that is obtainable from structure, and that this information is large in the solid phase of the model and decreases rapidly as state variables are varied into the fluid phase. 
} \\

\end{tabular}

 \end{@twocolumnfalse} \vspace{0.6cm}

  ]

\renewcommand*\rmdefault{bch}\normalfont\upshape
\rmfamily
\section*{}
\vspace{-1cm}

\footnotetext{\textit{$^{a}$~Department of Physics and Astronomy, University of Pennsylvania, 209 S. 33rd St, Philadelphia, PA 19104, USA. E-mail: itah@sas.upenn.edu, tsharp@sas.upenn.edu, ajliu@physics.upenn.edu}}
\footnotetext{\textit{$^{b}$~Department of Physics, Emory University, Atlanta, GA, USA. E-mail: daniel.m.sussman@emory.edu}}




\section{Introduction}
It has recently been discovered that the global dynamics of cells in dense epithelial tissues, are linked strongly to global structural properties. The dynamics of epithelial cells within some dense cultured monolayers slow to zero at the same time that the shapes of the cells approach a precisely-defined limit~\cite{park2015unjamming}. In a similar vein, the rate of cell dynamics correlates strongly with a relation between fluctuations in the aspect ratio and the mean aspect ratio of cells~\cite{atia2018geometric}. Even when such precise relationships do not hold, strong correlations can be found between the average shapes of cross-sectional views of cells and their dynamics in a variety of monolayers~\cite{devany2020cell}. Here we ask whether there is a link between {\it local} dynamics and {\it local} structure. Within dense tissue, movement of an individual cell must be accommodated by neighboring cells, and the possibility that the dynamics of a cell can be partially controlled by the geometry of its neighborhood could have implications for cell-cell diffusion, embryogenesis, or metastasis. 

The difficulty of finding a predictive connection between local structures and dynamics is well-known in the study of glass-forming materials. At the level of an individual particle, the question of whether the propensity of a particle to rearrange depends on its local structural environment has been addressed by many groups~\cite{tanaka2010,PhysRevLett.121.085703,doi:10.1063/1.3265983,PatrickRoyall2008,PhysRevLett.118.215501,chakrabarty2017block,Turciroyall,Leocmach2012}, and in recent years machine learning techniques have proven particularly powerful in identifying local structural quantities that predict dynamical events~\cite{cubuk2015,schoenholz2016structural,cubuk2016, schoenholz2016b,cubuk2017,sussman2017,sharp2018,ma2019heterogeneous,harrington2019machine,landes2020attractive, bapst2020unveiling,cubuk2020}. These methods allow for predictions of future local dynamics based on experimentally accessible imaging of the current state of the system.

A common approach in previous applications of machine learning to the study of disordered materials -- and the one we will follow in this work -- is to apply linear support vector machines~\cite{svm,scikit-learn} (SVMs) to first solve a classification problem in which representative examples of ``rearranging'' and ``non-rearranging'' particles are selected from molecular dynamics (or experimental!) trajectories. A linear combination of local structural features is sought that most strongly classifies these labeled datasets, represented by a classifying hyperplane in the high-dimensional space of local structural features. A scalar quantity, ``softness,'' is defined as the signed distance from a point in this high-dimensional space to the best-classifying hyperplane. In a variety of model glass-forming systems, this approach has uncovered a connection between softness and energy barriers to local rearrangement events: in other words, the propensity of a particle to undergo a rearrangement depends strongly on some linear combination of local structural features. Here we apply this machine learning methodology to show that there is a link between local cell dynamics and local cell geometry in numerical simulations of a model of dense 2D biological tissue. 

However, it is not enough to establish the existence of a link between local structure and dynamics  -- we also need to \emph{quantify} the strength of the link. Rather than focusing on the performance of our classifier when applied to training and test data sets, in this work  we introduce a quantification that compares the information gained about rearrangements given knowledge of cell softness with the maximum information that it is possible to gain from structural information. The latter quantity is calculated by performing many isoconfigurational simulations, i.e., by starting from the same configuration of cells in a tissue but drawing from independent realizations of the noise~\cite{Widmercooper2004isocon}. Remarkably, we will see that in some parts of model parameter space, linear SVMs come extremely close to capturing the maximum possible information about rearrangements that can be obtained from structure.

Our particular focus in this work is the Voronoi model (VM) of dense cellular matter, which captures important aspects of the \emph{cell-shape-dependent} nature of the rigidity transition in real epithelial tissues~\cite{bi2015density,park2015unjamming,devany2020cell}. The Voronoi model treats confluent cells as polygonal tilings governed by an energy functional that penalizes deviations of the perimeter and area of cells from preferred values. The target shape parameter $p_0$ -- the ratio of the preferred perimeter of a cell to the square-root of the area -- is a key variable that controls the mechanical response (rigidity transition) of the system. In the absence of self-propulsion, thermal noise, or some other means of generating cell rearrangements, the disordered state of this model is unambiguously rigid (has a positive shear modulus) for $p_0<p_c \approx 3.81$ and is mechanically unstable (has a vanishingly small shear modulus in the thermodynamic limit) for  $p_0 \gtrsim p_c$~\cite{bi2015density,sussmanmerkelnounjam2018}. Here we study both an equilibrium thermal version of the Voronoi model as well as an active version with self-propelled cells; at low temperature and activity scales, these model tissues exhibit some of the usual phenomena associated with nonlinear glassy dynamics, including caging, viscoelastic behavior, and dynamical heterogeneities~\cite{SussmanPaoluzziMarchetti2018}. 

There are several unusual features of the VM that make the study of it via the aforementioned machine-learning methods of immediate interest. From a biological perspective, previous work connecting ensemble averaged measures of local cell shape with overall system mobility suggests that the motile state of a monolayer can be inferred from static snapshots~\cite{bi2}; extending this inferential capacity to the probabilities that individual or small collections of cells will undergo coordinated motion is of obvious importance for a range of meso-scale properties of tissue dynamics . Methodologically, the Voronoi model is a highly anomalous glass-former, with \emph{sub}-Arrhenius scaling of the relaxation time with decreasing temperature~\cite{SussmanPaoluzziMarchetti2018}. 
While using linear SVMs to analyze softness works well in more ``well-behaved'' glassy systems, does it also work when the energy barriers apparently get \emph{smaller} as the temperature is lowered (hinting at the stress-controlled nature of the rigidity at $T=0$)? Does the apparent absence of quasi-localized low-frequency vibrational modes in $T=0$ packings of the VM indicate a fundamentally different connection between linear and non-linear response? Additionally, the VM has a parameter regime in which there is a vanishing shear modulus and apparently \emph{no} energy barriers to cellular motion~\cite{SussmanPaoluzziMarchetti2018,sussmanmerkelnounjam2018,sahu2020linear}. Does a methodology which correlates a learned feature (softness) with local energy barriers continue to work in a setting where such energy barriers vanish?

The remainder of this paper is organized as follows. In Sec. \ref{sec:methods} we outline both the numerical simulations of the thermal and active VM we studied, as well as the approach to calculating softness via linear SVMs. In Sec. \ref{sec:glassyResults} we apply this approach to study the behavior of softness in the ``solid-like'' regime of the model, i.e., for values of the target shape parameter $p_0<p_c$. In Sec. \ref{sec:fluidResults} we turn to the ``fluid-like'' regime of the model, and we show that our machine-learning methodology can correctly infer the nature of the model itself -- i.e., the SVM approach correctly identifies that characteristic energy barriers are softness-independent (and, in fact, vanish). We close in Sec. \ref{sec:disc} with a brief discussion.

%
%
\section{Methods}\label{sec:methods}
\subsection{Models studied}
 
We study the 2D Voronoi model~\cite{honda2004three,bi2,barton2017active} with the standard energy
functional,
\begin{equation}
\label{eq:vertex}
E = \sum^{N}_{i=1} \frac{1}{2}k_P (P_i - P_{0,i})^2 + \frac{1}{2}k_A (A_i - A_{0,i})^2.
\end{equation}
Here $P_i$ and $P_{0,i}$ are the actual and preferred perimeter of the cell $i$, and $A_i$ and $A_{0,i}$ are its actual and preferred area.
The first quadratic term models the effect of contractility of the acto-myosin cortex and an effective line tension arising due to cortical tension and cell-cell adhesion. The second quadratic term models a combination of volume  incompressibility of the 2D epithelial monolayer and its resistance to height fluctuations \cite{Hufnagel3835}. In this work we set the associated area and perimeter moduli $k_A$ and $k_P$ to unity. The degrees of freedom of this model are cellular positions, and the cellular geometry is determined via a Voronoi tessellation of the set of cellular positions. We simulate this space-filling model of dense tissue in a periodic domain, choosing the linear size and length units such $L = \sqrt{\sum_i A_{0,i}} = \sqrt{N}$. We focus on monodisperse systems, i.e., systems in which all cells are assigned identical values of $A_{0,i}$ and $P_{0,i}$, but note that we find qualitatively similar results when studying bidisperse systems as well. In our monodisperse system, the dimensionless ratio $p_0\equiv P_0 / \sqrt{\langle A \rangle}$ defines the target shape parameter which controls the mechanical and dynamical state of the tissue~\cite{merkel2018geometrically,teomy2018confluent}.

We use the open source ``cellGPU'' package \cite{Sussman2017cellGPUMP} to simulate both thermal and self-propelled versions of the overdamped Voronoi model. In both cases, the cell positions, $\vec{r}_i$, are updated according to the equation of motion,
\begin{equation}
    \frac{d \vec{r}_i}{dt}=\mu \vec{F}_i + \vec{\eta}_{i}
\label{EOM}
\end{equation}
where the constant $\mu$ is the inverse friction coefficient and $\vec{F}_i \equiv - dE/d\vec{r}_i$ is the conservative force on cell $i$.
In the thermal model at temperature $T$, the stochastic term
models random forces. Its components are uncorrelated white noise, with zero mean and $\langle \eta_{i\alpha}(t)\eta_{j\beta}(t^\prime)\rangle =2 \mu T \delta(t-t^\prime)\delta_{ij}\delta_{\alpha \beta}$, where $\alpha$ and $\beta$ are the  Cartesian components.

In the self-propelled model, each cell experiences time-correlated random forces and the stochastic term is instead $\vec{\eta}_i = v_0 \hat{\eta_{i}}$. The constant $v_0$ sets the magnitude of the motility and $\hat{\eta_{i}}=(cos (\theta_i), sin (\theta_i))$ is the polarization vector of each cell. The director $\hat{\eta_{i}}$ rotates randomly according to
$\partial_t\theta_i=\xi_i(t)$, where again, $\xi_i(t)$ is uncorrelated. $\langle \xi_i(t)\xi_j(t^\prime)\rangle =2D_r\delta(t-t^\prime)\delta_{ij}$
with the rotational diffusion constant $D_r$ characterizing the scale of the orientational noise. For the self-propelled model, we define an effective temperature~\cite{PhysRevLett.108.235702} $T_{eff}=\frac{v_0^2}{2\mu D_r}$. This mapping to an effective temperature is not without controversy -- particularly in systems where the degrees of freedom are not free to diffusive around but rather may be caged to local positions. Nevertheless, it is a valuable guide to one's intuition, and one expects the effective temperature mapping to be exact~\cite{C3SM52469H} in the limit where rotational diffusion constant $D_r \to \infty$ at fixed ``effective inertia'' $(\mu D_r)^{-1}$ .

We study both thermal and active versions of the Voronoi model to check the robustness of our machine learning model and validate various forms of cellular motion governed by different equations of motion. Indeed, the precise mechanism of cell movement is itself an important research area, and if the machine-learning approach we advocate here is to be applied to real experimental systems it is important to establish the degree to which the physical interpretation of our results depend on the precise microscopic details.

As a first step towards this, here we establish a baseline for which the equilibrium and active models are expected to give similar results. To this end, we show results for the active model at $\mu = 1.0$ and $D_r=50.0$ except where noted. Most of our simulations were carried out with time steps ranging from $\Delta t = 0.01$ to $\Delta t = 0.1$, depending on the temperature, self-propelled velocity, and/or rotational diffusion constant under consideration -- in all cases we confirmed that our structural and dynamical results are insensitive to simulating with an integration time step an order of magnitude smaller in size. 

For a wide array of $p_0$ values and $T$ or $T_{eff}$ values, we performed $100$ independent simulations of $1024$ cells each. We did initial thermalization at each target (effective) temperature for at least $10^6$ time steps before recording data. In studying the low-temperature fluid phase of the model in Sec. \ref{sec:fluidResults} we additionally performed sets of independent simulations of $N=5000$-cell systems, initializing for $10^7$ time steps and recording data over  $5\times 10^6$ subsequent time steps.

\subsection{Characterization of local dynamics}
 In atomic, molecular, or colloidal glasses one typically must choose a threshold value of some dynamic indicator, for instance $D^2_{min}$ or $p_{hop}$ ~\cite{falk1998dynamics,PhysRevLett.102.088001}, in defining a characteristic rearrangement or cage-breaking event. In contrast, in the space-filling models of dense cellular matter considered here, rearrangements can be naturally defined by changes in the set of cells that a target cell shares an edge with. For the majority of this paper we focus on these changes in the set of nearest neighbors for a given cell as the signature of a rearrangement event. We neglect neighbor changes associated with cell division or death events~\cite{czajkowski2019glassy}, and only permit neighbor exchanges via T1 transitions, an event in which an edge separating two cells vanishes and an edge separating two of their neighbors appears. Note that the T1 event marks the transition of the system from one metastable minimum to another in the energy landscape, and the initial length of the edge that disappears is correlated with the energy difference between the two metastable minima~\cite{kimHilgenfeldt2018universal, popovic2020inferring}.

Because the length of an edge that disappears during a T1 event must vanish during the event, small edge lengths are themselves predictive of future rearrangement events~\cite{popovic2020inferring}. However, edge length is only one of many quantities that correlate with T1 events, and we show here that a linear combination of structural quantities, which could be readily extracted not only from simulation snapshots but from video microscopy experiments on real dense tissue, combine to give a much stronger predictor of the propensity of a cell to rearrange than the minimum edge length.

\subsection{Calculation of softness}
To find a structural quantity that correlates strongly with T1 events, we first construct separate training sets (to build our classifier on) and test sets (to validate our learned features and evaluate the quality of our predictions). For Sec.~\ref{sec:glassyResults}, we build a labeled training set by scanning through long molecular dynamics trajectories of the Voronoi model conducted at a particular point in parameter space (i.e., for a particular value of $p_0$ and $T$), and identify representatives from two characteristic populations of cells: those that will participate in a T1 rearrangement event within a short future time window, and those that will not participate in a T1 event for a longer future time window (``rearranging'' and ``non-rearranging'' cells, respectively). The rearranging training set consists of $N_{r}=4000$ cells that participated in a T1 event within $4\tau$ of the time we identify them, and the non-rearranging training set consists of $N_{nr}=4000$ cells that do not rearrange during at least a $160\tau$ time window. As in previous work on softness~\cite{schoenholz2016structural}, the construction of a combined training set using multiple thresholds to select out both rearranging \emph{and} non-rearranging cells is important to maximize the effectiveness of this approach.

For each cell of the combined training set, we identify a large collection of quantifiers of the local structural environment. We have used three different sets of structural variables to construct the SVM. The first set is the standard one used in Refs.~~\cite{cubuk2015,schoenholz2016structural,cubuk2016,schoenholz2016b,cubuk2017,sussman2017,sharp2018,landes2020attractive}, introduced for machine learning applications by Behler and Parrinello~\cite{BehlerParinello2007}. The second set consists of additional measures of the cellular geometry, such as those those listed in Table~\ref{correlationwithS} (the full list is provided in the Appendix.) The results here are for the third set, which is the union of the two sets of structural quantities, but prediction accuracy results are quantitatively nearly identical for each of the sets, indicating that there is considerable redundancy between the two sets. Indeed, the spirit of our effort to classify local structure in some high-dimensional space does not require that we look for optimal or orthogonal sets of structural descriptors. For the rearranging cells, we calculate these structural features on a configuration before the T1 event occurs, and for the non-rearranging cells we calculate these structural features in the center of the time window over which they do not rearrange. We emphasize that in none of the three sets of structural features we consider do we include the minimum edge length of the cell and its neighboring cells.

We standardize all structural features, so that they all have zero mean and unit variance for the training set. Given this set of standardized local structural features matched to labels for rearranging and non-rearranging cells, we use a linear support vector machine to find the optimal hyperplane separating the classes in the training set. The softness for cell $i$, $S_i$, is defined as the signed distance of the point corresponding to cell $i$ (defined by the values of the structural variables for cell $i$) to the hyperplane. Cells with points on the rearranging side of the hyperplane have $S>0$ while those with points on the non-rearranging side have $S<0$.

\section{Softness in the solid-like regime}\label{sec:glassyResults}
We first apply the above methodology to the Voronoi model in the regime of target shape parameters in which the model would be unambiguously rigid at zero temperature ($p_0<p_c$). The scaling of the relaxation time in this regime is sub-Arrhenius, but the low-temperature behavior can still be thought of as governed by effective energy barriers to T1 transitions~\cite{sahu2020linear}. Specifically, then, we train a classifier using data from simulations at $p_0=3.75$ and $T=2.0\times 10^{-3}$; we then \emph{apply} this learned classifier to data at other temperatures and also at other \emph{effective} temperatures in the case of the self-propelled model. 

In this section, we will first show that, when applied to new data at the same set of parameters we trained at, softness accurately classifies cells according to their propensity to execute a T1 transition. To help quantify how well softness succeeds at this task, we measure the information gained about T1's through softness. We find that the information gained does a very good job of reflecting the glassy dynamics of the tissue, and the information gained decreases as both $p_0$ and $T$ increase, i.e., as the relaxation time in the model decreases and motion becomes more fluid-like. We then show that, as in all other other glass-formers studied so far~\cite{schoenholz2016structural,ma2019heterogeneous,cubuk2020}, softness correlates with an effective energy barrier. Remarkably, even the sub-Arrhenius dynamics in the VM are accurately predicted by shifts in the average value of softness as the temperature is varied.

\subsection{Properties of softness and cross-validation accuracy}
\begin{figure}[]
\centering
  \includegraphics[width=0.45\textwidth]{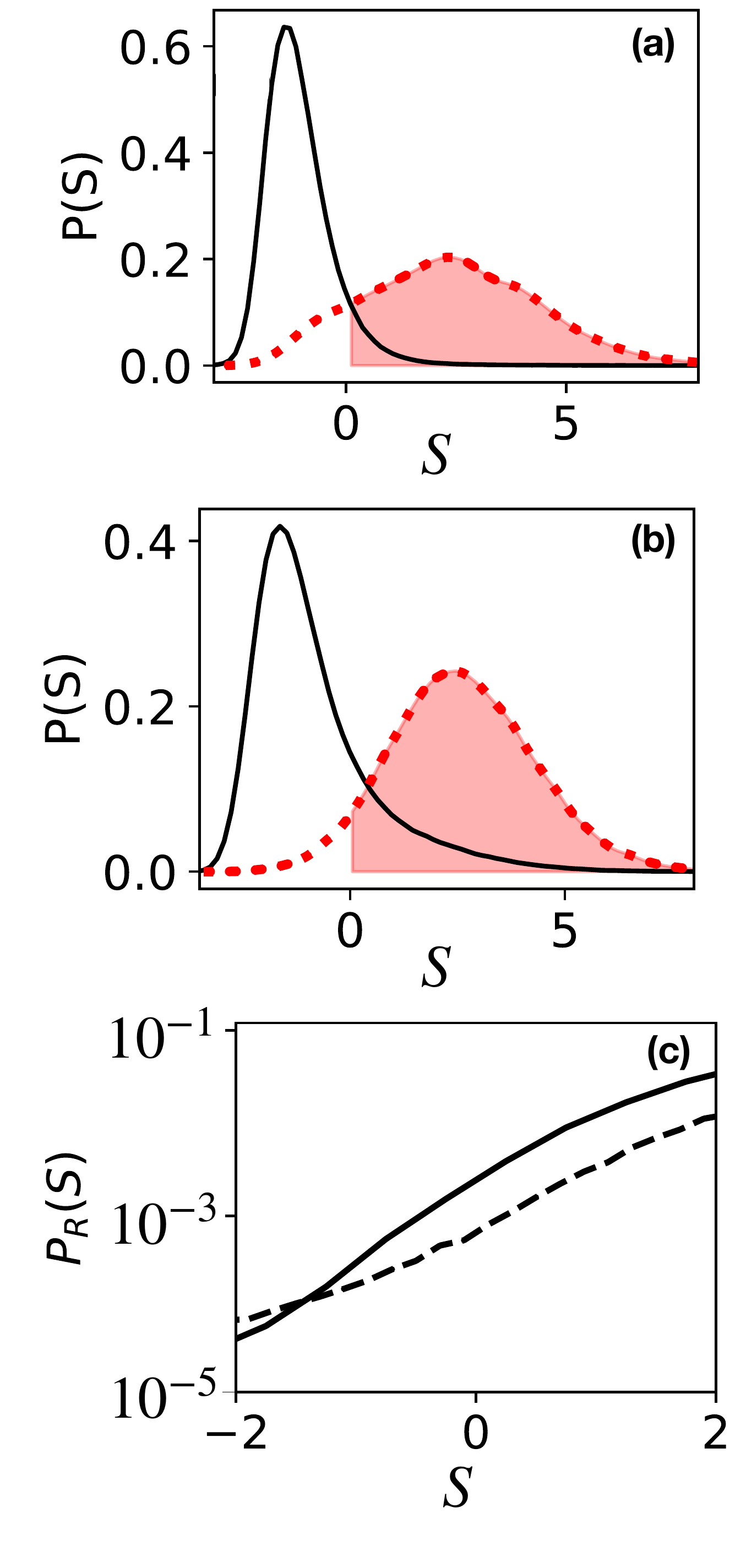}
  \caption{Probability distribution of softness for all cells (black solid curve) and for cells that are about to
rearrange (red dashed curve) for (a) the thermal Voronoi model at temperature $T = 1.5 \times 10^{-3}$ and target shape parameter $p_0=3.75$, and (b) the self propelled Voronoi model at propulsion speed 
$V_0 = 0.5$ and $p_0=3.75$. (c) 
The probability of rearrangement for cells  as a function of their softness
  value for the thermal system (solid) and self-propelled systems (dashed).}
  \label{CVM_softness}
\end{figure}

After training our classifier, we apply it to a large quantity of previously unseen simulation data to measure distributions of softness and the probability that a cell will rearrange as a function of the assigned softness value. The distribution of cell softness, $P(S)$, is shown in black at $p_0=3.75$ for the thermal system at $T=1.5\times 10^{-3}$ (Fig.~\ref{CVM_softness}(a)) and the self-propelled system at $V_0=0.5$ and $D_r=50$ (Fig.~\ref{CVM_softness}(b)).  In the same panels we plot the distribution of softness for cells that are about to rearrange, $P(S|R)$, in red; as initial indications of reasonable generalization, we find that the two distributions, $P(S)$ and $P(S|R)$, are well separated and that roughly $90\%$ of the cells about to rearrange have $S >0$.

One way to assess the capacity of softness to predict rearrangements is to calculate the cross-validation accuracy (CV). To calculate the  ``ten-fold" CV accuracy, for example, a training set is divided into 10 equal parts. Nine of the ten parts are used to train and obtain a hyperplane, and a classification accuracy is calculated on the remaining 10\% of the data, called the test subset (where the classification accuracy is simply is defined as the fraction of cells in the rearranging test subset with $S>0$). This process is repeated 10 times in total, once for each choice of test subset; the average prediction accuracy over the 10 trials is the CV accuracy. The CV accuracy is useful for preventing an over-fitting of the test set and thereby over-estimating the accuracy of the classifier.
We have found that the CV accuracy varies from $79 \%$ to $92 \%$ depending on temperature and self propulsion speed, with lower accuracy at higher $T$ and $V_0$. This is comparable to values obtained from fitting a wide variety of disordered solids and glassy liquids~\cite{cubuk2017}.

We emphasize that the CV accuracy we report is a crude measure of the predictive capacity of our learned classifier. It concerns the binary classification of just two subsets of continuously distributed dynamical events, and as such depends on the thresholds used. Our interest, though, is not the maximum CV accuracy, but the physical interpretation of $S$. A more complete and more physical way of assessing the success of $S$ is to calculate $P_R(S)$, the fraction of cells of a given $S$ that are about to rearrange. In Fig.~\ref{CVM_softness} (c) we show the probability of rearrangement $P_R$ as a function of softness $S$ for the thermal and self-propelled systems, respectively. Evidently, $P_R(S)$ varies by 2-3 orders of magnitude with changing $S$. This result shows that, as in previous studies of softness, the scalar value of softness correlates directly and strongly with the probability that a cell is about to rearrange.

\begin{figure}
\centerline{
\includegraphics[width=0.50\textwidth]{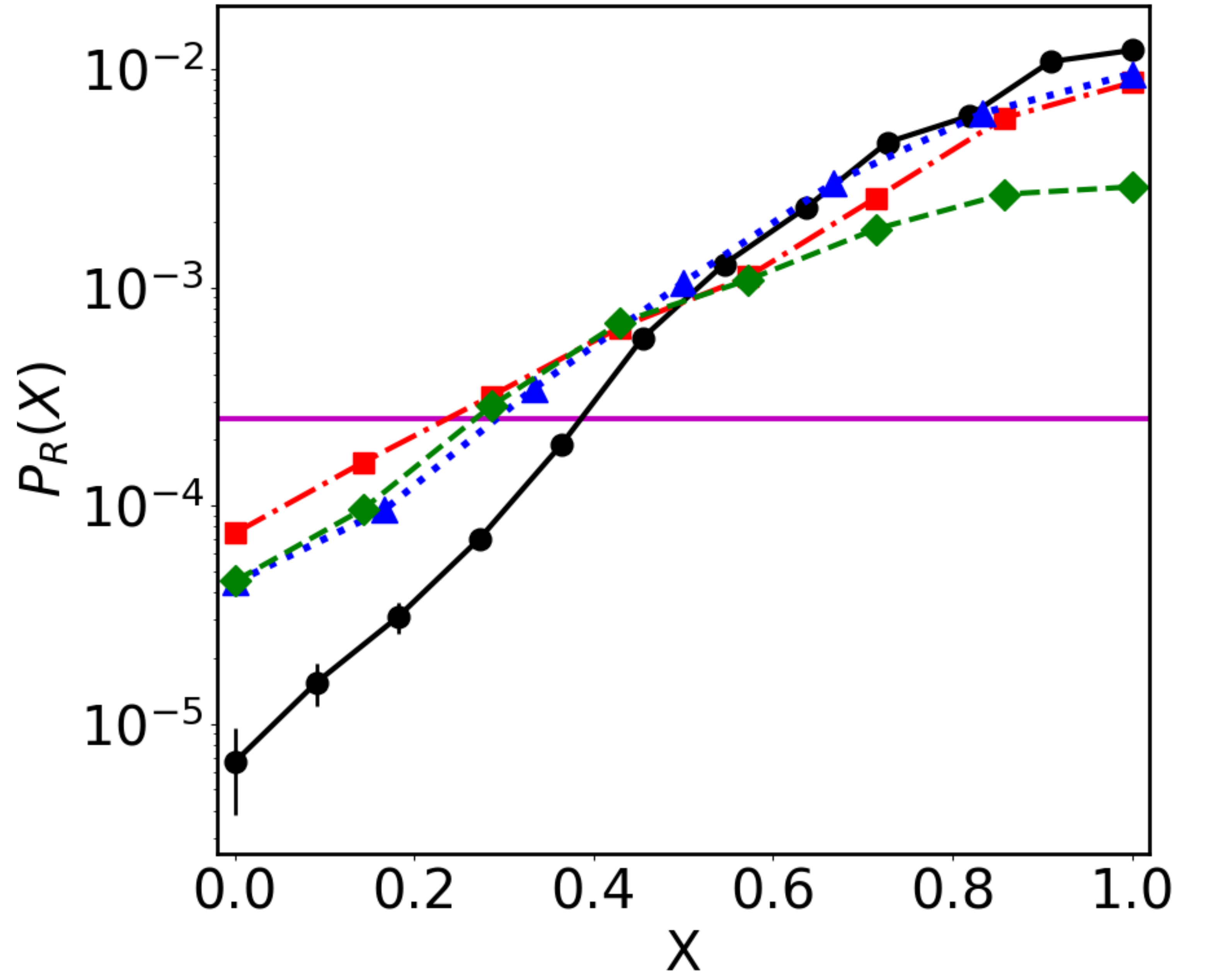}
}
\caption{\label{Qpercentile}
Probability to rearrange vs the percentile $X$ in the thermal system of each structural quantity: softness--black circles connected with solid lines, cell shape parameter $p$--red squares, dash-dotted lines, maximum shape parameter of cell neighbor $(p_\text{hn})$ - blue triangles, dotted lines, minimum length of cell edge $1/l_\text{min}$ - green diamonds, dashed lines. Of all the quantities, softness gives the strongest variation of $P_R(X)$. The magenta horizontal line represents the average rearrangement rate for cells. (Same state as Fig.~\ref{CVM_softness}.)
}
\end{figure}

The predictiveness of softness may be compared with that of other local structural quantities. In Fig.~\ref{Qpercentile}, $P_R$ is again plotted, but for several different measures of local structure. All of these quantities are mapped to the interval $[0,1]$ by plotting $X$, the percentile rank in different structural quantities. The values of $P_R$ for different percentiles are shown also for the cell's softness (black circles), the cell's shape parameter, $p \equiv P / \sqrt{\langle A \rangle}$ (red squares),  the highest shape parameter among the cell's neighbors, 
$p_\text{hn}$ (blue triangles), and the length of the shortest edge of the cell, $l_\text{min}$ (green diamonds). For $l_\text{min}$, both the cell's own edges and those edges next to the cell are considered, since a T1 transition that eliminates any of those implies a change of neighbors. 

A quantity that yields a curve that lies on the average rearrangement rate (horizontal magenta line), would completely fail to predict rearrangements, since the cells would rearrange at the average rate irrespective of their value of that quantity. In order for the quantity to be predictive, its $P_R$ curve should deviate strongly from the average rearrangement line. Clearly, the range of $P_R$ spanned by softness $(S)$ (black solid line) is higher than for any other structural quantity studied. This means that the probability of undergoing a rearrangement is more sensitive to softness. How can this sensitivity be quantified, and how can that quantification be compared in an absolute sense to the total information contained in the local structure, independent of our ability to measure it?

\subsection{Quantifying information gained}
To quantify how well softness predicts rearrangements, we are first led to consider the general question, how much information does structure contain about dynamics? Performing multiple simulations starting from the same set of particle positions but drawing independent realizations of their velocities, in what is termed the isoconfigurational ensemble~\cite{Widmercooper2004isocon}, provides a powerful tool for answering this question. It is straightforward to apply this idea to quantify the short-time propensity of every cell in a given configuration. For each cell $i$, one can calculate the fraction of all realizations in which the cell participated in a T1 event during a given short time window, $P_R^{\text{iso}}(i)$.

We use these ideas, building on previous work that constructed length scales in disordered materials to via mutual information~\cite{dunleavy2015mutual}, and quantify the information gained by knowing the softness of a cell by proceeding as follows. In a fixed short time interval in a thermal system, there are two possible outcomes for each cell -- the cell does or does not participate in a T1 transition. The Shannon entropy associated with this process is 
$$H(p) = - p \text{log}_2 p - (1-p) \text{log}_2 (1-p),$$
where $p$ is the probability that the cell participates in a T1 event.

Let $p_i$ be the probability that a given cell $i$ participates in a T1 event. Suppose all cells have the same probability to rearrange, so that $p_i = \langle P_R \rangle$. The entropy per cell is maximal this case: $H_\text{max} \equiv H(\langle P_R \rangle)$. Any cell-specific knowledge about the probability to rearrange will decrease this entropy, down to a limit of zero in the hypothetical scenario where the outcome for every cell is predicted perfectly.

Knowledge of a structural quantity like softness modifies the entropy associated with a cell whose softness is $S$ via our (measured) knowledge of $P_R(S)$. The decrease in entropy corresponds to an information gain; averaged over the entire system the information gained per cell is
\begin{equation}
I_S = \sum_i (H_\text{max} - H(P_R(S_i))) /N.
\end{equation}
At the other extreme, if a quantity $Z$ is unrelated to rearrangements then $P_R(Z) = \langle P_R \rangle$ and $I_Z = 0$; i.e., no information about rearrangements is gained by knowing $Z$. This corresponds to the horizontal magenta line on Fig.~\ref{Qpercentile}. Of all quantities investigated, $S$ (black solid line) deviates most from the horizontal line and consequently contains the most information about rearrangements.

How close does softness come to the limit imposed by the system itself? 
Softness provides an \emph{estimate} of the true probability for a cell to rearrange; we calculate the true probability of rearrangement for each cell by performing many isoconfigurational simulations, all starting from an identical configuration of cells but with random noise realizations in Eq.~\ref{EOM}. The maximum information that can be gained from structural information, according to our isoconfigurational calculation, is
\begin{equation} 
I_\text{iso} = \sum_i (H_\text{max} - H(P_R^{\text{iso}}(i))) /N.
\end{equation}
In Table 1 we report the information gained by softness and compare it to this isoconfigurational calculation of the maximum possible information gain. We also compare to the other most predictive features we have identified, and to other simple geometric features that have historically been considered.

\begin{table}[h]
 \label{tbl:example}
  \begin{tabular}{|c| c | c | c | c | c | c |}
    \hline
        $\{p_0,T\}$ & $\frac{I_{\textrm{iso}}}{H_\text{max}}$ & $\frac{I_{S}}{I_{\textrm{iso}}}$ & $\frac{I_{p}}{I_{\textrm{iso}}}$ & $\frac{I_{nhp}}{I_{\textrm{iso}}}$  & $\frac{I_{l_{min}}}{I_{\textrm{iso}}}$ &  $H_\text{max}$  \\
    \hline  
    $\{3.75,10^{-2.82}\}$  & 0.202 & 0.889 & 0.500 & 0.556  & 0.806&  0.018 \\ 
    $\{3.75,10^{-2.45}\}$   & 0.116 & 0.857 & 0.098 & 0.537 & 0.642&  0.0524\\ 
    \hline
  \end{tabular}
\small
  \caption{Bits of Information per cell encoded in various structural features. The subscripts $iso$, $S$, $p$, $nhp$, $l_{min}$ denote iso-configuration, softness, the shape index of the cell itself, the highest shape index of any neighboring cell, and the cell minimum edge length. $H_\text{max}$ is the information that would be gained by perfect prediction.
  }
\end{table}

Although no quantity we measure gains the maximum available information, we see that softness captures a demonstrably large fraction of the total information that structure can provide, as measured by $I_\textrm{iso}$. Other structural measures capture a much smaller amount of information, with only the length of the shortest edge coming close to the information gained via softness (which we again emphasize was not considered as a structural features in building the softness classifier). We note that these results partially support the intuition that small edge lengths are predictive of future rearrangements, but that combining information about the shortest  edges  with many other structural features gives a more robust  predictor for future cell rearrangements throughout the parameter space of the model.

\subsection{Information gained from softness at different points in the phase diagram}
\begin{figure}[!h]
\centering
  \includegraphics[width=0.45\textwidth]{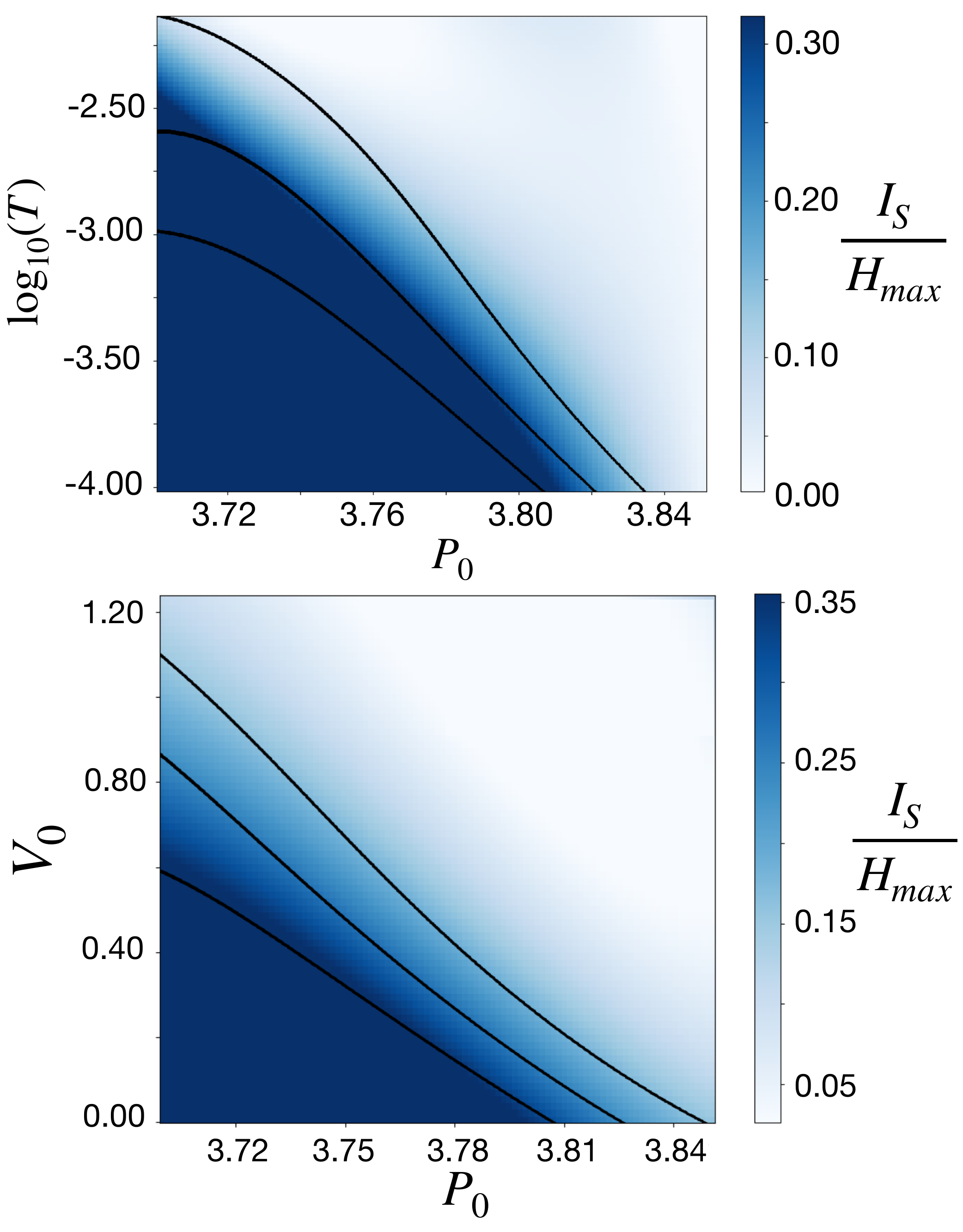}
  \caption{Top Panel: The fraction of information extracted from the local structure (bluescale) approximately follows the decrease in the rate of T1 transitions: lines of constant $\langle P_R \rangle$ ($\langle P_R \rangle = 10^{-5}, 10^{-3.8}, 10^{-2.8}$ from left to right) show the increase in rearrangement rate with temperature for thermal system. Bottom panel: Similar plot for self-propelled system where lines of constant $\langle P_R \rangle$ ($\langle P_R \rangle = 10^{-3.5}, 10^{-3.0}, 10^{-2.5}$ from left to right) show the increase in rearrangement rate with propulsion speed for self-propelled system.} 
  \label{phasediagram}
\end{figure}

We have seen that $I_S$ provides a reasonable approximation of the maximum information at a particular state point, and it has the virtue that it can be readily calculated directly on our large ensemble of trajectories without the need to compute expensive isoconfigurational simulations associated with every simulation snapshot.
The ratio $I_S/H_{\text{max}}$ gives a measure of both the importance of local structure and also how heterogeneously the rearrangement dynamics in a system are distributed. We report this in Fig.~\ref{phasediagram}, which shows that this measure depends strongly on ($p_0$, $T$) for thermal systems and 
on ($p_0$, $v_0$) for self-propelled systems.
At larger $p_0$ and $T$ (or $T_{eff}$), the softness $S$ contains vanishingly small amounts of information about whether T1 transitions will occur, as one might expect, although we note  that even  in  the ``fluid''-like regime of the model local structural information has \emph{some} predictive power. This suggests that the typical paths cells explore as they perform T1 transitions still involves going over small-but-finite energy barriers.
A contour of constant relaxation rate (black line), $\langle P_R \rangle$, is shown for comparison. The plot is remarkably similar to those obtained previously for thermal and self-propelled systems when plotting the average shape $\langle p \rangle$ and diffusion constant in the $p_0-T$ or $p_0-v_0$ planes~\cite{SussmanPaoluzziMarchetti2018,bi2}. The agreement between the contours of constant $\langle P_R \rangle$ and contours of shades of blue shows that the average rearrangement rate is highly correlated with the amount of information about rearrangements captured by softness, with more information from softness at lower rearrangement rates. This suggests that the amount of information captured by softness depends strongly on the average rearrangement rate.

\begin{figure}[]
\centering
\hskip 0.2in
  \includegraphics[width=0.43\textwidth]{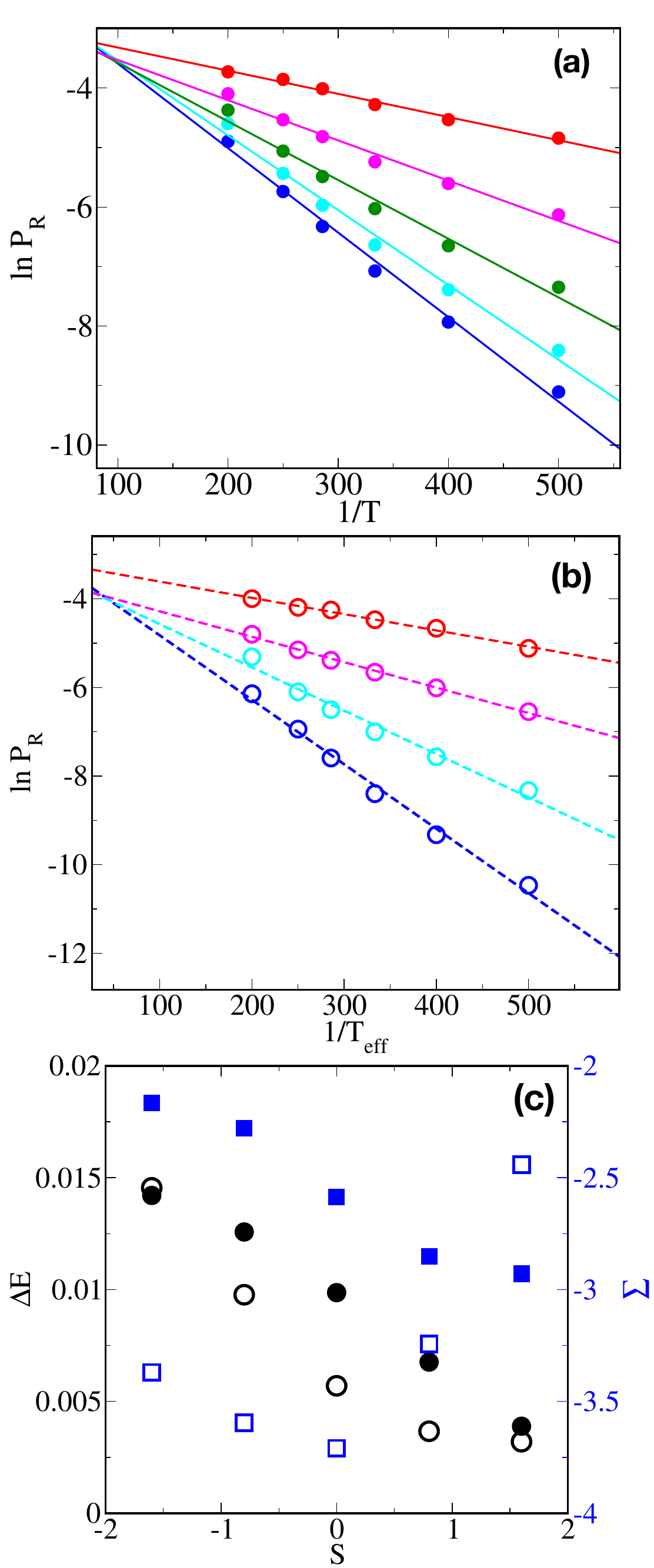}
  \caption{$P_R(S)$ as a function of  $1/T$ for different softness values ranging from $S \sim -2$ (blue) to $S \sim 2$ (red) for (a) the thermal system and (b) the self-propelled system. For the latter system we use $T_{eff}$ (see text). (c) The quantities $\Sigma(S)$ (blue squares) and $\Delta E(S)$ (black circles) defined in Eq.~\ref{eq:arr} for the two systems, with data for the thermal system indicated by solid symbols and data for the self-propelled system indicated by open symbols.}
  \label{Arrhenius}
\end{figure}

\subsection{Arrhenius behaviour of $P_R(S)$}
The top panel of Fig.~\ref{Arrhenius} shows $\ln P_R(S)$ as a function of $\frac{1}{T}$. An Arrhenius relation, corresponding to a well-defined barrier to rearrangement, corresponds to a straight line on this plot.
We find the probability of rearrangement for cells of a given softness $S$ follows an Arrhenius relationship with $1/T$, \begin{equation}\label{eq:arr}
P_R(S) = \exp(\Sigma(S)-\frac{\Delta E(S)}{T})
\end{equation}
 where the configurational ($\Sigma(S)$) and and energy barrier ($\Delta E(S)$) terms do not themselves depend on temperature $T$. Similar results have been found in other systems~\cite{schoenholz2016structural,schoenholz2016b,sussman2017,sharp2018,ma2019heterogeneous,cubuk2020}. The result also holds for $T_{eff}$ in the self-propelled particle system. Note that $P_R(S)$ becomes nearly independent of $S$ (the Arrhenius curves for different softness cross) at an ``onset'' temperature near the left axis in the figure. This is consistent with the results of previous studies on supercooled liquids: in Lennard-Jones systems above the onset temperature, $T_0$, structure does not predict rearrangements ~\cite{schoenholz2016structural}.  In the bottom panel of Fig.~\ref{Arrhenius} we show the dependence of $\Sigma$ and $\Delta E$ on softness for the thermal and self-propelled systems. 
These results are consistent with the interpretation that below the onset temperature  cells that have large negative softness values are in local configurations with higher energy barriers to rearrangements, whereas cells that have large positive softness values are not.
 The variation of these quantities with $S$, together with the distribution of softness itself, supports the finding that at low (effective) temperatures there are broad distributions of dynamically heterogeneous regions of the system~\cite{SussmanPaoluzziMarchetti2018}.

\subsection{Temperature-dependence of the softness distribution, and the relationship between structure and dynamics}
The Arrhenius behavior of $P_R(S)$ may come as a surprise, particularly as the behavior of the relaxation time itself in these systems is sub-Arrhenius: it suggests that softness is providing a decomposition of the system into locally Arrhenius components, and that the sub-Arrhenius behavior of the system overall stems from the way the distribution $P(S)$ shifts to lower values of $S$ with decreasing temperature ~\cite{schoenholz2016structural,landes2020attractive,cubuk2016}. 
In supercooled liquids with super-Arrhenius behavior it was found that the the overall super-Arrhenius scaling of the relaxation time in those systems was, indeed,  consistent with the magnitude of this overall shift in the softness distribution together with the detailed probability of rearranging at different values of softness. In the Kob-Andersen model, this careful balance came together in the statement that the average softness of a system controls its structural relaxation time~\cite{schoenholz2016b}
\begin{equation}
\label{eq:tau}
\tau_\alpha \sim \frac{1}{P_R(\langle S \rangle)}.
\end{equation}
To reiterate: in those earlier studies a classifier was trained at a single temperature, the probability of rearrangement as a function of softness was shown to obey Eq.~\ref{eq:arr}, and the classifier was applied to new simulation data at different temperatures to obtain $\langle S \rangle$ at each $T$. From this information they accurately predicted the structural relaxation time as a function of $T$.

\begin{figure}[h]
\centering
  \includegraphics[width=0.47\textwidth]{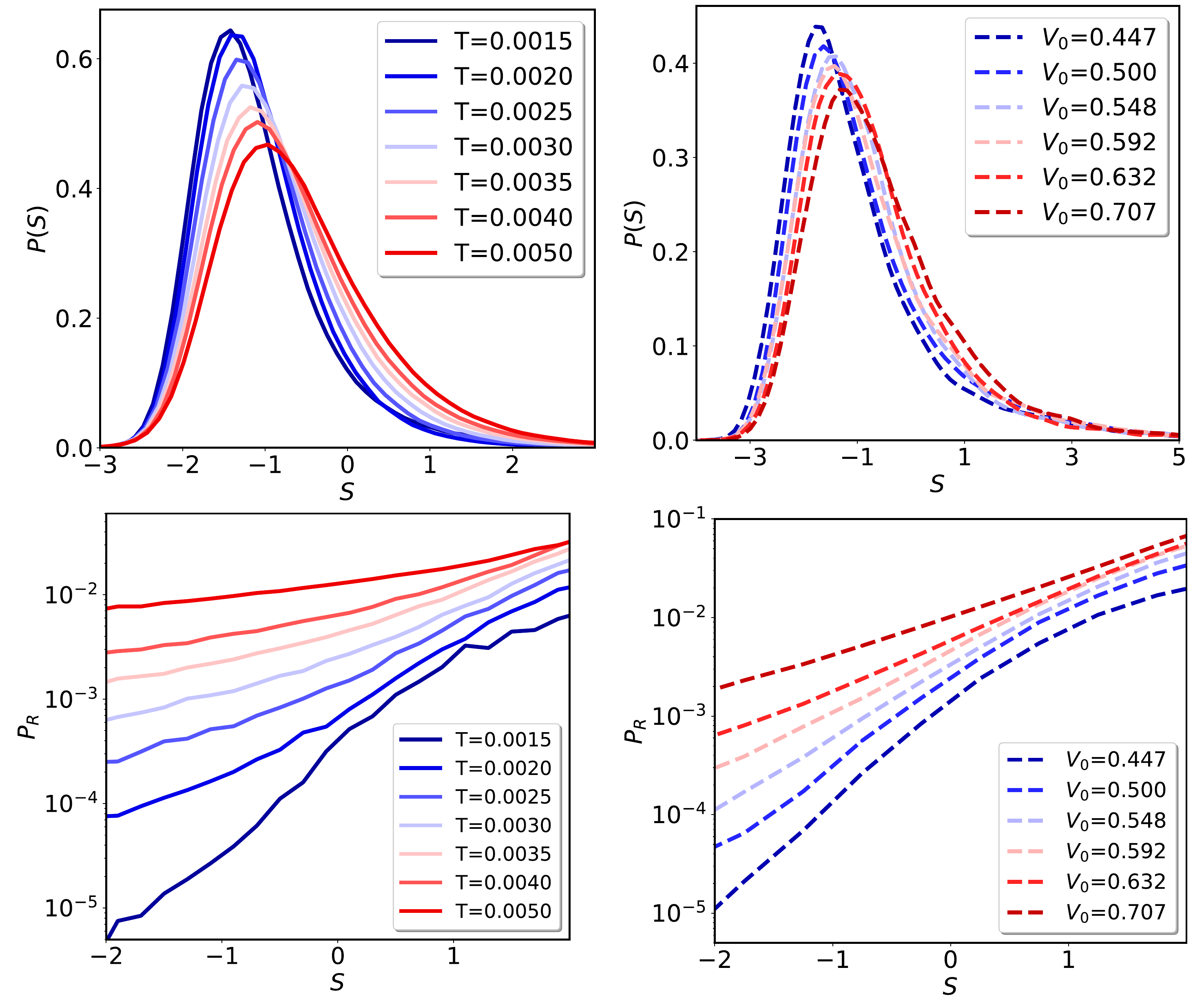}
  \caption{The softness distribution for (a) the thermal system at different temperatures $T$ and (b) the self-propelled system at different propulsion speeds $V_0$. The probability of rearrangement $P_R(S)$ for cells as a function of their softness value for (c) thermal systems at different $T$ and (d) self-propelled systems at different $V_0$.}
  \label{PS_vs_T}
\end{figure}

How does this story play out in the case of a model with \emph{sub}-Arrhenius scaling of the dynamics? The top panel of Fig.~\ref{PS_vs_T} shows that for the Voronoi models, the distribution of softness similarly shifts slightly to lower values of $S$ with decreasing temperature and self propulsion speed. In the bottom panel of Fig.~\ref{PS_vs_T} we show the probability of rearrangement $P_R(S)$ as a function of cell softness $(S)$ value for the thermal system and for the self-propelled system.

\begin{figure}[]
 \centering
  \includegraphics[width=0.40\textwidth]{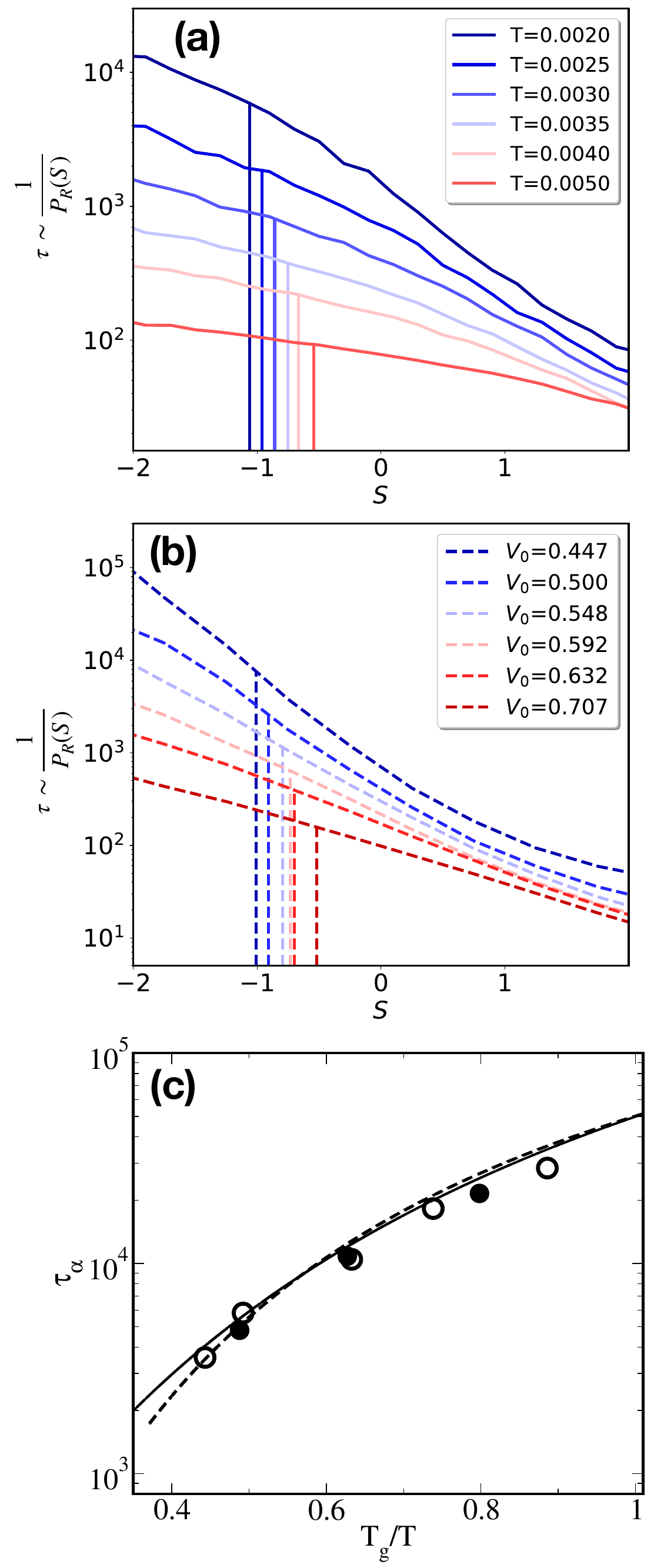}
  \caption{The inverse probability of rearrangement $1/P_R(S)$ as a function of softness $S$ for (a) the thermal Voronoi model and (b) the self-propelled model. Vertical lines represent the equilibrium  average softness $\langle S \rangle$, and their upper limits indicate the relaxation time $\tau_\alpha$ predicted by Eq.~\ref{eq:tau} at each temperature $T$ and self propulsion speed $V_0$.
(c) Angell plot of relaxation time (defined as $Q(\tau_\alpha) = \exp{(-1)}$) for thermal (solid circles) and self-propelled (open circles) Voronoi models. The solid and dashed lines are the relaxation times predicted by Eq.~\ref{eq:tau} for the thermal and self-propelled models, respectively.  For the self-propelled model, we define an effective temperature $T_{eff}=\frac{V_0^2}{2\mu D_r}$. Temperature is normalized by the value of $T_g$ where relaxation time is $\tau_\alpha = 5 \times 10^4$.
}
  \label{StrDyn}
\end{figure}

In Fig.~\ref{StrDyn} (a) and (b) we display the relaxation time $(\tau \sim 1/(P_R(S))$ as a function of  softness $S$ for our model systems. The vertical lines represent the mean softness of the system, with color indicating the value at each temperature or propulsion speed. Surprisingly, here, too, the combination of shifts in the average value of softness as the temperature of a system is changed and the probability of rearranging at a given value of softness conspire to accurately reproduce the temperature dependence of the relaxation time. This is shown in Fig.~\ref{StrDyn}(c), where we plot the relaxation time as measured by the dynamics together with that predicted by Eq.~\ref{eq:tau}.

\section{Properties of softness in the fluid-like regime}\label{sec:fluidResults}
In the previous section we applied the methodology of Ref.~\cite{schoenholz2016structural} -- with appropriate modifications for a model in which a threshold-free definition of rearrangement events can be defined in terms of T1 events -- to study a system in which sub-Arrhenius scaling of the relaxation time is observed and which possesses a nonzero shear modulus at zero temperature. We found that, as in previous studies of more standard glass-forming systems, a predictive classifier can be built, that the relaxation time of the system is well-characterized by $\tau_\alpha \sim \frac{1}{P_R(\langle S \rangle)}$, and that $S$ itself has a physical interpretation as a local energy barrier scale to rearrangement.

What happens, then, when we apply the SVM approach to predict dynamics from local structure in a parameter regime of the model, $p_0 \gtrsim 3.81$, in which the shear modulus is very small at zero temperature~\cite{sussmanmerkelnounjam2018}, there are only vanishingly small energy barriers to nonlinear cell motions, and the relaxation time scales as approximately a simple power law over the entire range of temperatures studied\cite{SussmanPaoluzziMarchetti2018}? Figure \ref{phasediagram} indicated a small amount of information about T1 transitions was gained even in the fluid regime, suggesting that the cells are traversing small but finite energy barriers as they performing T1 transitions. However, this is likely because at the relatively high temperatures studied above, the model is off the manifold of zero-energy motions. Is it possible to modify the details of how we trained the classifier in order to correctly identify the existence of barrier-free paths to relaxation?

To examine these questions, we train a classifier that is different from the one studied in the previous section. The previous classifier was trained at $p_0=3.75$ and $T=2.0 \times 10^{-3}$. Here we train a classifier at $p_0=3.85$ and $T=9.7\times10^{-4}$ (a parameter-space point for which the structural relaxation time is approximately $\tau_\alpha = 10^4$). We then apply this classifier to additional simulation data at much lower temperatures than those studied in the previous section, at $T=\{3\times10^{-6},\ 6\times10^{-6},\  1\times10^{-5},\ 2.9\times10^{-5},\ 5.8\times10^{-5},\ 9.7\times10^{-5},\ 2.9\times10^{-4},\ 5.8\times10^{-4},\ 9.7\times10^{-4} \}$. While low temperatures in absolute terms, this range of temperature corresponds approximately to alpha relaxation times stretching from $10^4$ to $10^7$. 

For these very cold systems in the fluid regime, T1 transitions are very rare and it is difficult to build sufficiently large training sets from them or to obtain reliable estimates of the very low probability of a cell experiencing a T1 transition. We therefore return to the use of a ``hop'' indicator function, $P_h(i;t)$ \cite{PhysRevLett.102.088001}, to identify dynamical events: to define $P_h(i;t)$ for a cell $i$ at time $t$ we first specify two time intervals $A=\left[ t-5\tau,t\right]$ and $B=\left[t, t+5\tau\right]$; the hop indicator function can then be expressed as
\begin{equation}
P_{h}(i;t)=\sqrt{  \langle \left( \mathbf{r}_i - \langle \mathbf{r}_i\rangle_B\right)^2\rangle_A \langle \left( \mathbf{r}_i - \langle \mathbf{r}_i\rangle_A\right)^2\rangle_B  },
\end{equation}
where $\langle\rangle_A$ and $\langle \rangle_B$ denote averages over $A$ and $B$ intervals.

\begin{figure}[h]
\centering
  \includegraphics[width=0.45\textwidth]{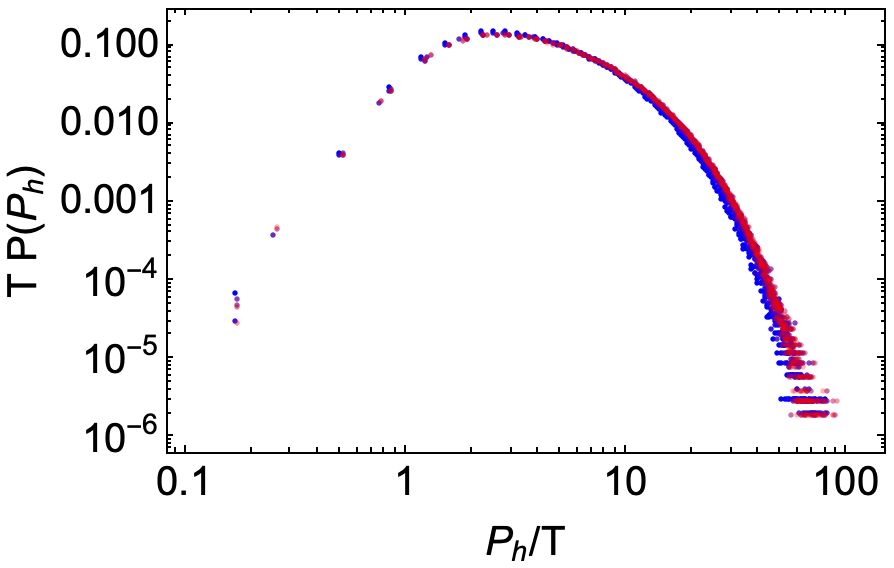}
    \includegraphics[width=0.45\textwidth]{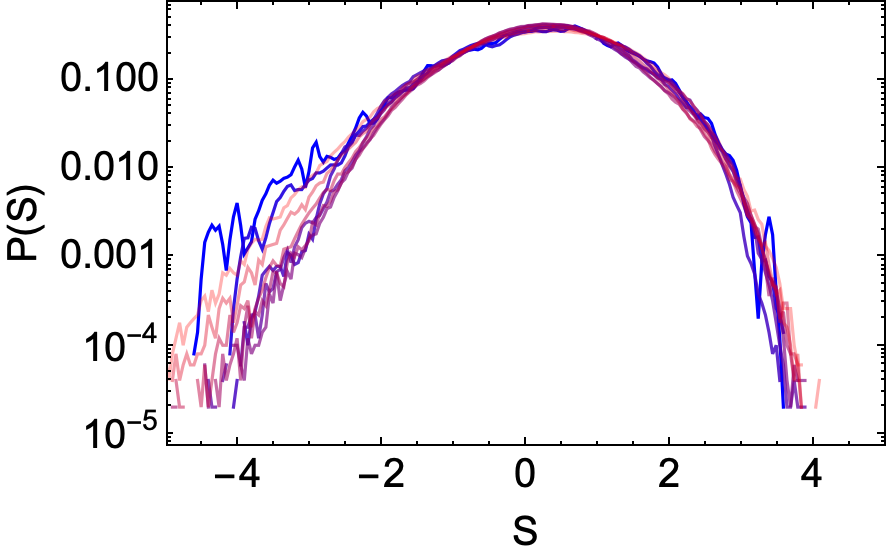}
  \caption{(Top) Probability distribution of $P_{h}$, scaled by temperature, for the thermal VM at $p_0 = 3.85$  and $T=3\times10^{-6}  -  9.7\times10^{-4}$ (blue to red points). (Bottom) The log of the probability distribution of softness for the same systems, measured using a classifier trained at $T= 9.7\times10^{-4}$. The distribution is nearly unchanged, despite the temperature (and the relaxation time) changing by over two orders of magnitude.}
  \label{fluidSoft}
\end{figure}

The probability distribution of this indicator function is reported in the top panel of Fig. \ref{fluidSoft}. Notably, for these low temperatures and in a system where the dynamical timescale itself simply grows as $1/T$, the distribution of values of the indicator function collapses perfectly when scaled by the temperature -- this temperature dependence is also what one would expect in studying the distribution of the $P_h$ indicator as applied to any Brownian random walk. This behavior is consistent with power-law growth of the overall relaxation time at this state point. Note, though, that in more usual particulate glasses (e.g., binary Lennard-Jones mixtures) the probability of rearrangements changes by orders of magnitude with decreasing temperatures, but this simple collapse of the $P_h$ distribution would not occur. Instead, when rare rearrangements do occur in those systems, the statistics of particle motions is quantitatively similar to rearrangements occurring at other temperatures.

We follow the protocol outlined in Ref.~\cite{schoenholz2016structural}, choosing a cutoff value of $P_h$ that selects only the top $\sim 5\%$ of all $P_h$ values as an ``active'' cell, defining ``inactive'' cells as those that stay below another a small threshold of $P_h$ for several hundred $\tau$ to ultimately form a balanced set of cells in our training set. We train a linear SVM -- using the same Behler and Parrinello~\cite{BehlerParinello2007} structure functions (specialized to the monodisperse case) as were used in Ref.~\cite{schoenholz2016structural} --  and, consistent with the low amount of information gained by softness in this regime (as seen in the previous section), our cross-validation accuracy is comparatively low: we only identify cells participating in dynamical events in our training set with $\sim 70\%$ accuracy. The distributions of softness themselves across our studied temperature regime are shown in the bottom panel of Fig. \ref{fluidSoft}; as can be seen, other than a small change in the low-softness tail, there is almost no change in this distribution across a broad range of temperature scales, quite unlike the results in the previous section.

The next step in the standard analysis is to calculate $P_R(S)$ at different temperatures. The collapse Fig.~\ref{fluidSoft} indicates that if we use a single threshold for $P_h$ to determine what a ``rearrangement" is, there will be no events at sufficiently low temperatures; already-rare events at $T= 9.7\times10^{-4}$ would be essentially unobservable at $T=3\times10^{-6}$ within the limitations of our simulations.  But the collapse also suggest a different way to define events. Reflecting the fact that the low-temperature dynamical events in this model are themselves very different, we make use of the data collapse in Fig.~\ref{fluidSoft} and define an active cell as having \emph{equally extreme values of $P_h$ at any given temperature}; that is, we scale the threshold we use to define a rearrangement by the temperature (rather than using a temperature-independent, fixed threshold as in previous work).

By looking systematically at the probability of ``rearranging'' with this definition of dynamical events and comparing with Eq.~\ref{eq:arr}, we can still decompose $P_R(S)$ into a configurational part $\Sigma(S)$ and an energy barrier scale $\Delta E(S)$ across all our data covering two orders of magnitude in temperature. We find $\Delta E(S) = 0$ to within numerical accuracy. That is, by training in this regime not only do we (correctly) find that local structure does not indicate an energy barrier scale, we also find that there is also no temperature-dependent part of $P_R$ at all -- i.e., the softness analysis indicates that there are no energy barriers at low temperature, exactly as expected from the nature of the model~\cite{sussmanmerkelnounjam2018,SussmanPaoluzziMarchetti2018}.

We repeat this analysis to study parameter space points at $p_0=\{3.75,\ 3.8,\ 3.85,\ 3.9\}$ and over comparably lower temperature regimes. In each case, we use the classifier trained at $p_0=3.85$. We find that $\Delta E=0$ for $p_0 = 3.85$ and $p_0 = 3.9$, and that it is \emph{nonzero} for $p_0 = 3.75$ and $p_0 = 3.8$. Apparently the combination of features that maximizes the (limited) predictive capacity of local structural information at these very low temperatures in the fluid phase nevertheless contains sufficient information to predict the emergence of nonzero energy barriers in the solid phase. We find this result striking. It likely has strong implications for the reason why linear SVMs give softnesses that are interpretable as energy barrier scales in the first place, and we are  continuing to study this intriguing result.

\section{Discussion}\label{sec:disc}
We have shown that local structure in these tissue models suffices to obtain a quantity, softness, that strongly predicts rearrangements, and have quantified the amount of information that it captures. The information provided by softness is strongly correlated with the relaxation time (contours of constant information  shape parameter are nearly the same as contours of constant average T1 rate). The results of the present study, when combined with previous work on both strong and fragile glass-formers, raise several natural questions about the ``softness'' protocol for identifying local structures that are predictive of activated dynamics. Most pressingly, under what conditions does softness correctly identify an energy barrier scale in the problem? And to what extent does the answer to that question depend on the nature of the model under study and to the details of how the training sets are constructed?

Remarkably, we find that a single classifier trained in a low-temperature but fluid-like regime of the Voronoi model correctly predicts both the absence of energy barriers in the fluid-like regime and the presence of them in the solid-like regime. It may be useful to contrast this finding with expectations in a ``normal'' system going from a simple fluid to supercooled fluid / glassy phase. In the normal fluid phase there are not meaningful barriers to particle rearrangements, nor are there dynamical heterogeneities of the sort seen at lower temperatures. After building a classifier to attempt to identify fluid particles in the tails of the Gaussian distribution of particle displacements by looking at local structures, one would not expect this same classifier to correctly identify local structures that appropriately describe dynamical heterogeneities in the glassy phase. Nor would one expect, even if this classifier had some predictive accuracy, it would necessary be physically interpretable as encoding the energy barriers to local rearrangements.
 
In glassy systems, the lowest-frequency modes are quasilocalized; this reflects the physics that the corresponding rearrangements are localized. In contrast, the lowest-frequency vibrational modes of the ground states of the Voronoi model are spatially \emph{extended}~\cite{SussmanPaoluzziMarchetti2018}, initially suggesting that rearrangements or at least the motions associated with rearrangements are extended. Given this, it may be surprising that local structure succeeds in predicting rearrangements so well. This is related to the fascinating difference in the linear vs. the non-linear mechanics of the low-temperature states of the VM. This difference is also reflected in the previous finding that, despite the extended linear vibrational modes of the VM found by quenching to zero temperature, the spatial extent of dynamical heterogeneities (as inferred from the value of four-point correlation functions) is similar to those in more typical glass-forming models~\cite{SussmanPaoluzziMarchetti2018}. Thus, while the low-frequency vibrational modes are extended, the physics governing nonlinear motions rearrangements may not be.

The fact that softness, which predicts T1 events, reproduces the temperature-dependence of the structural relaxation time in the glassy regime of the model indicates that the T1 transitions are central to the dynamics within the tissue. Softness also suggests that there is a spectrum of effective energy barriers in a tissue, which accounts for orders of magnitude variation in the rate of T1s. It would be interesting to apply our learned classifier to an even more disparate set of data than we have done above. In addition to more thoroughly testing our classifiers on very different model points (e.g., on values of the preferred shape parameter even farther from the training value than we have reported here), one could imagine studying even greater perturbations of the model, such as the addition of intrinsic curvature~\cite{PhysRevResearch.2.023417}.

Our results for self-propelled models represent, we believe, the first analysis of softness in an active system. In this paper we have restricted ourselves to the regime in which the self-propelled system is reasonably well-described by an effective temperature, so it is relatively unsurprising that our methodology continues to work. The analysis approach is not limited to that regime, however, and it will be interesting to study systems far from that limit. As the non-equilibrium nature of the active simulations are enhanced, how different do the statistics of the energy barriers to rearrangements become? At what point does maintaining good predictive capacity require the use of additional structural features, such as the relative polarization directions of nearby cells? Answering these questions will be a first step towards applying this style of analysis \emph{directly} to experimental datasets on epithelial tissues and cell cultures. In those that behave much like Voronoi and vertex models, we could even ask whether the learned classifier from this \emph{model} of dense 2D epithelial tissue is able to maintain any predictive capacity when applied to experimental data, or whether additional features (cell polarization, levels of protein expression within cells, etc.) are necessary to predict cell rearrangements in real epithelial tissues.

\subsection*{Conflicts of interest}
There are no conflicts of interest to declare.

\subsection*{Acknowledgements}
We thank J. Fredberg, M. L. Manning and S. S. Schoenholz for instructive discussions. This project was primarily supported by the National Cancer Institute of the National Institutes of Health under Physical Sciences Oncology Center (PSOC, physics.cancer.gov) award No. U54 CA193417 (TAS, IT and AJL). Additional support was provided by the National Science Foundation under grant DMR-1506625 (TAS, IT) and the Simons Foundation for the collaboration Cracking the Glass Problem via Award 454945 and Investigator Award 327939 (AJL).

\subsection*{Appendix}

We use two sets of structure functions to quantify the local environment around each cell for input to the support vector machine. The first set consists of those commonly used~\cite{BehlerParinello2007} that depend directly on the configuration of cell centers. We augment these  with a second set of structural features that depend directly on the cell shapes, obtained from the Voronoi tessellation.

Comprising the first set, each radial structure function $R_{i\alpha}$ indicates the number of neighbors at a particular distance from the cell $i$.

\begin{equation*}
R {(i;\alpha)} = \sum_j \text{exp}(-(d_{ij} - \mu_{\alpha})^2/(2\sigma_0^2))
\text{,\hspace{5mm}}
\end{equation*}

The vector from the position of cell $i$ to the position of cell $j$ has length $d_{ij}$. 
We include 24 radial functions.
We use $\sigma_0=0.1$ and $\mu_\alpha = 0.7,0.8,...,2.9,3.0$ corresponding to $\alpha = \{ 0,1 \ldots 13,14,15 \ldots 22,23 \}$;

In the second set, the following 10 values are included for each cell $i$: the number of Voronoi edges of cell $i$, then the cell own shape index $p_i$, and the shape index of the cell's most compact neighbors, $p_j$ for $j=0...7$,
where  $p_j$ is the shape index of the Voronoi neighbor with the $j-th$ smallest shape index (or zero if cell $i$ has less than $j$ Voronoi neighbors).
With these 10 Voronoi-dependent values, in total we consider $M=34$ structure functions.

For rearranging cells, structural functions are calculated between 4 to $8\tau$ before a T1 transition; this way, the the structure reflects the local tissue that may rearrange, rather than when it is actively rearranging. For non-rearranging cells, the structure is calculated in the middle of the 160-$\tau$ window.

Given this, we write our training set on $N$ examples as $\{(\textbf{F}_1,y_1),(\textbf{F}_2,y_2),..,(\textbf{F}_N,y_N)\}$. Here $\textbf{F}_i = \{F_i^1,F_i^2,..,F_i^M\}$ is the set of $M$ input structural features (discussed above) that describes the local neighbourhood of cell $i$, and $y_i$ is the label for example $i$. We use the label $y_i = 1$ for rearranging cells, and $y_1 = -1$ for non-rearranging cells. Then, we use the SVM \cite{svm,scikit-learn} algorithm with a linear kernel to find the optimal hyperplane  $\vec{W} \cdot \vec{F} -b =0$ that separates the rearranging cells $(y_i=1)$ from the non-rearranging cells $(y_i=-1)$. 
$\vec{W}$ is the vector of weights associated with each structural feature, and $b$ is the bias.

We find a classifying hyperplane for a training set obtained from a single temperature and preferred perimeter, and apply that learned hyperplane to the rest of our data to obtain the results reported above: given any cell $k$ we find its local structural features,  $(\textbf{F}_k)$, and define a scalar continuous variable, ``softness,'' as $S_n = \vec{W} \cdot \textbf{F}_k -b $, which is signed distance of the point $\textbf{F}_k$ to the hyperplane.

\subsection*{Which structural variables dominate softness}
One way to determine which structural variables for softness are most important is to use recursive feature elimination~\cite{rfe}. Here, we instead look at which structural variables have the strongest correlations with $S$. We calculate Pearson correlation coefficients with softness for each of the structural variables (after normalization as discussed in the SI) used in SVM. The correlation coefficients  are listed in Table~\ref{correlationwithS}.  Of all quantities calculated, the neighbor which have largest shape index is most highly correlated with softness. The cell's own shape index is essentially equally well-correlated with $S$, as is the highest shape index of the neighbor cells. 

An alternative analysis based on the accuracy of an SVM in predicting T1's if trained only on one of these structural features found similar results.

\begin{table}[h]
\centering
\small
  \caption{\ Pearson correlation coefficient ($C$) of local quantities with $S$ ($p=3.75, T=1.5 \times 10^{-3}$). Negative correlations are indicated by (-). ``Neighbors near $d=1.5$'' refers to the count of cells between distance $d-0.05$ and $d+0.05$ from the cell, \emph{i.e.} in a radial bin.}
  \label{correlationwithS}
  \begin{tabular*}{0.35\textwidth}{@{\extracolsep{\fill}}r|l}
    \hline
    $C$ & Feature \\
    \hline

0.75     & Neighbor largest shape index \\
0.63     & Mean shape index of neighbors \\
0.61     & Cell own shape index \\
0.59     & Number of neighbors near d=1.5 \\
0.46     & Number of neighbors near d=1.6 \\
0.46     & Number of neighbors near d=2.5 \\
0.42     & Number of neighbors near d=2.4\\
-0.34     & Number of neighbors near d=1.8\\
-0.32     & Number of neighbors near d=2.8\\
-0.32     & Number of neighbors near d=2.9\\
0.32     & Number of neighbors near d=2.6\\
-0.28     & Number of neighbors near d=1.1\\
0.25     & Number of neighbors near d=0.8\\
0.23     & Number of neighbors near d=0.9\\
-0.22     & Number of neighbors near d=1.2\\
-0.12     & Number of neighbors near d=1.0\\
-0.12     & Number of neighbors near d=2.0\\
-0.07     & Number of neighbors near d=2.1\\
-0.05     & Number of neighbors near d=2.2\\
0.04    & Number of neighbors \\
    \hline
  \end{tabular*}
\end{table}



\balance



\bibliography{rsc} 
\bibliographystyle{rsc} 

\end{document}